\documentclass[twocolumn,aps,prl,10pt]{revtex4-1}
\usepackage{amsmath}
\usepackage{graphics}
\usepackage{epsfig}
\usepackage{verbatim}   
\usepackage{color}
\usepackage{setspace}

\begin{document}

\title{Coherent Control at Its Most Fundamental: Carrier-Envelope-Phase-Dependent
\newline
Electron Localization in Photodissociation of a H$_2^+$ Molecular Ion Beam Target}
\author{T. Rathje$^{1,2}$}
\author{A.~M.~Sayler$^{1,2}$}
\author{S.~Zeng$^{3}$}
\author{P.~Wustelt$^{1}$}
\author{H.~Figger$^{4}$}
\author{B.~D.~Esry$^3$}
\author{G. G. Paulus$^{1,2}$}
\email{gerhard.paulus@uni-jena.de}
\affiliation{$^1$Institute for Optics and Quantum Electronics, Friedrich Schiller University Jena, Max-Wien-Platz 1,07743 Jena, Germany}
\affiliation{$^2$Helmholtz-Institut Jena, Helmholtzweg 4, D-07743 Jena, Germany}
\affiliation{$^3$J.R. Macdonald Laboratory, Kansas State University, Manhattan, Kansas, 66505 USA}
\affiliation{$^4$Max Planck Institute of Quantum Optics, Hans-Kopfermann-Strasse 1, 85748 Garching, Germany}
\begin{abstract}
Measurements and calculations of the absolute carrier-envelope-phase (CEP) effects in the photodissociation of the simplest molecule, $\rm{H}^{+}_{2}$, with a 4.5-fs Ti:sapphire laser pulse at intensities up to $(4\pm2)\times10^{14}$\,W/cm$^2$ are presented. Localization of the electron with respect to the two nuclei (during the dissociation process) is controlled via the CEP of the ultrashort laser pulses. In contrast to previous CEP-dependent experiments with neutral molecules, the dissociation of the molecular ions is not preceded by a photoionization process, which strongly influences the CEP dependence. Kinematically complete data are obtained by time- and position-resolved coincidence detection. The phase dependence is determined by a single-shot phase measurement correlated to the detection of the dissociation fragments. The experimental results show quantitative agreement with \textit{ab inito} 3D time-dependent Schr\"odinger equation calculations that include nuclear vibration and rotation.
\end{abstract}

\maketitle

Chemical reactions are governed by the dynamics of electrons. Using light to move electrons around within molecules in order to coherently control the dynamics of chemical reactions therefore appears to be the ultimate approach. Although experiments with complex laser pulse shapes and pump-probe schemes have shown that this is feasible (see e.g. Ref. \cite{Assion1998}), understanding and interpreting the results is often challenging. An alternative approach that has been used to control attosecond dynamics in various strong-field processes \cite{Christov1997,Paulus2003,Kruger2011,Xie2012} is the manipulation of the absolute carrier-envelope phase, $\phi$, of few-cycle laser pulses, $E(t)$=$E_{0}(t) \cos(\omega t+\phi)$, with the pulse envelope $E_{0}(t)$ and frequency $\omega$.

Here we demonstrate coherent control of electron localization and the fragmentation rate in the simplest molecule, H$_2^+$ (D$_2^+$), with essentially a single optical cycle, by manipulating the field evolution with the absolute phase. In contrast to previous experiments \cite{Xie2012,Kling2006a,Kremer2009,Sansone2010}, control of the electron is not due to laser-induced ionization dynamics and is thus more relevant as a benchmark for photochemical reactions. To realize this simplest possible scenario of coherent control, we start directly from H$_2^+$ and implement a kinematically complete measurement.

So far, electron localization with intense few-cycle laser pulses has only been explored starting from neutral molecules. Even in the simplest case (H$_2$ and isotopologues), these are multielectron systems and, as such, not yet amenable to accurate theoretical treatment. Electron localization in H$_2$ is thought to proceed in a multistep scenario, see, e.g., Refs. \cite{Kling2006a, Kremer2009, Znakovskaya2012}. An initial ionization step promotes the nuclear wave packet to the H$_2^+$ $1s\sigma_{g}$ potential \cite{Posthumus2004}, followed by dissociation  caused by (i) recollision with the first electron \cite{Niikura2002} or (ii) dissociation of the promoted nuclear wave packet. These two mechanisms can be mostly -- but not completely -- separated in the kinetic energy release (KER) spectrum. It should be noted that when the absolute phase is used to control the dissociation of H$_2$, control of electron localization is highly correlated with the well-known strong phase dependence of the ionized recolliding electron \cite{Milovsevic2003}.
 
Another paradigm deserving experimental verification, which is tested here, is the question of to what extent can H$_2$ measurements be treated with H$_2^+$ models?  When starting from H$_2$, the H$_2^+$ nuclear wave packet is a coherent superposition of vibrational states in the $1s\sigma_{g}$ electronic ground state, in contrast to the incoherent Franck-Condon distribution of vibrational levels present when starting from H$_2^+$ produced in an ion source \cite{Amitay1999}. Furthermore, the prerequisite ionization step of the two-electron system H$_2$ precludes modeling the creation of the initial H$_2^+$ without assumptions \cite{Kelkensberg2011}, very much in contrast to the one-electron H$_2^+$ system, where \textit{ab initio} calculations can be done \cite{Anis2008}.

\begin{figure*}
\begin{center}
\includegraphics[width=0.9\textwidth,angle=0]{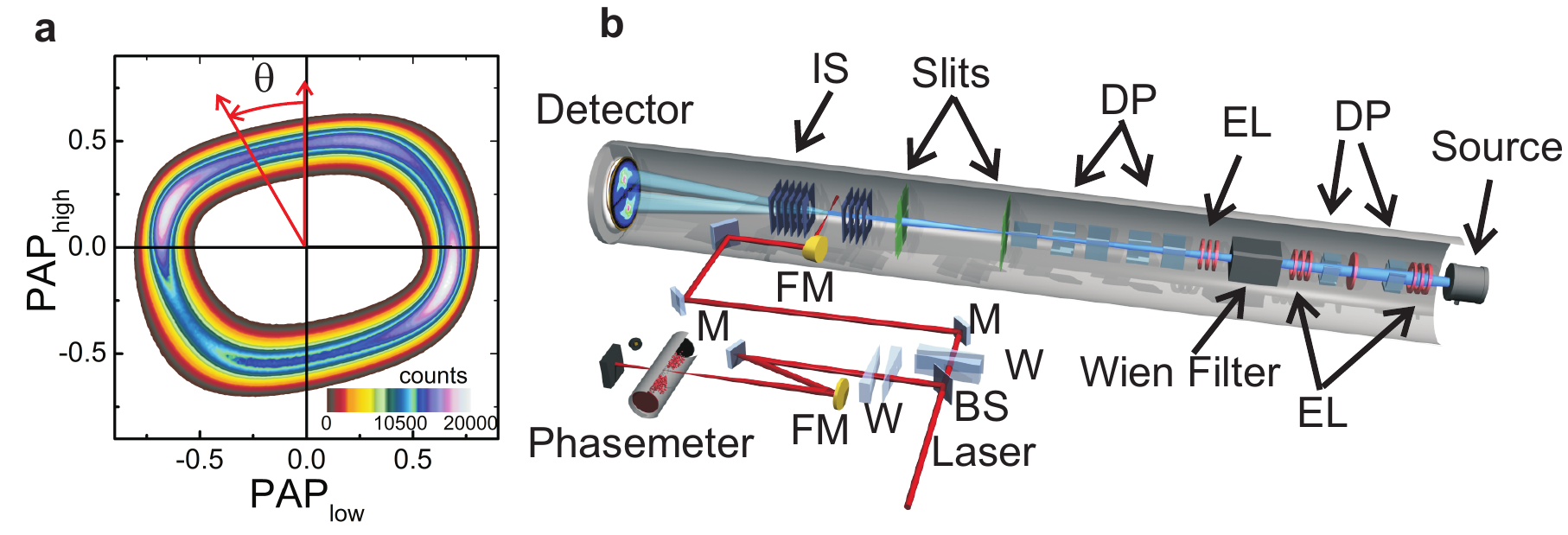}
\end{center}
 \caption{\textbf{a}, The parametric phase asymmetry plot (PAP) measured by the phasemeter. The angle $\theta$ corresponds to the absolute phase $\phi$. $\theta$ is recorded for each and every laser pulse with a single-shot error of less than 100\,mrad (for details, see Ref. \cite{Rathje2012b}). \textbf{b}, Schematic drawing of the ion beam and phasemeter setup. The ion beam apparatus employs a time and position sensitive microchannel plate detector with a delay line anode to record every reaction fragment in coincidence and synchronized with the phasemeter signals. (DP, deflector plates; EL, einzel lense; IS, ion spectrometer; M, silver mirror; BS, beam splitter; W, glass wedges; FM, focusing mirror. For more details see the Supplemental Material.)}
 \label{fig:exp_setup}
\end{figure*}

Although several measurements of the laser-induced fragmentation of H$_2^+$ have been done, e.g. Refs. \cite{Sandig2000, BenItzak2005, McKenna2012}, absolute phase-dependent measurements starting from the molecular ion, e.g., H$_2^+$, have been absent. This is due to the technical difficulty associated with the realization of a setup providing sufficient stability for a number of sensitive parameters such as the absolute phase over extended (tens of hours) data acquisition times. We have overcome these difficulties using a novel phase tagging technique (see Fig.~\ref{fig:exp_setup}) that has made phase stabilization obsolete for a large class of experiments \cite{Rathje2012b, Bergues2012}. The decisive new aspect of this work is that the nuclear degrees of freedom -- and therefore the position of the electron quantified by the asymmetry of the electron distribution -- are controlled via the interaction of a H$_2^+$ molecular ion beam target with a 4.5\,fs few-cycle laser pulse with a center wavelength of 700\,nm and a peak intensity of $(4\pm2)\times10^{14}$\,W/cm$^{2}$. The ion target is generated within a ion beam setup. The whole setup is used for photodissociation of H$_2^+$, while the full 3D momenta of the fragments are measured in coincidence and as a function of the absolute phase.

\begin{figure}
\begin{center}
\includegraphics[width=0.45\textwidth,angle=0]{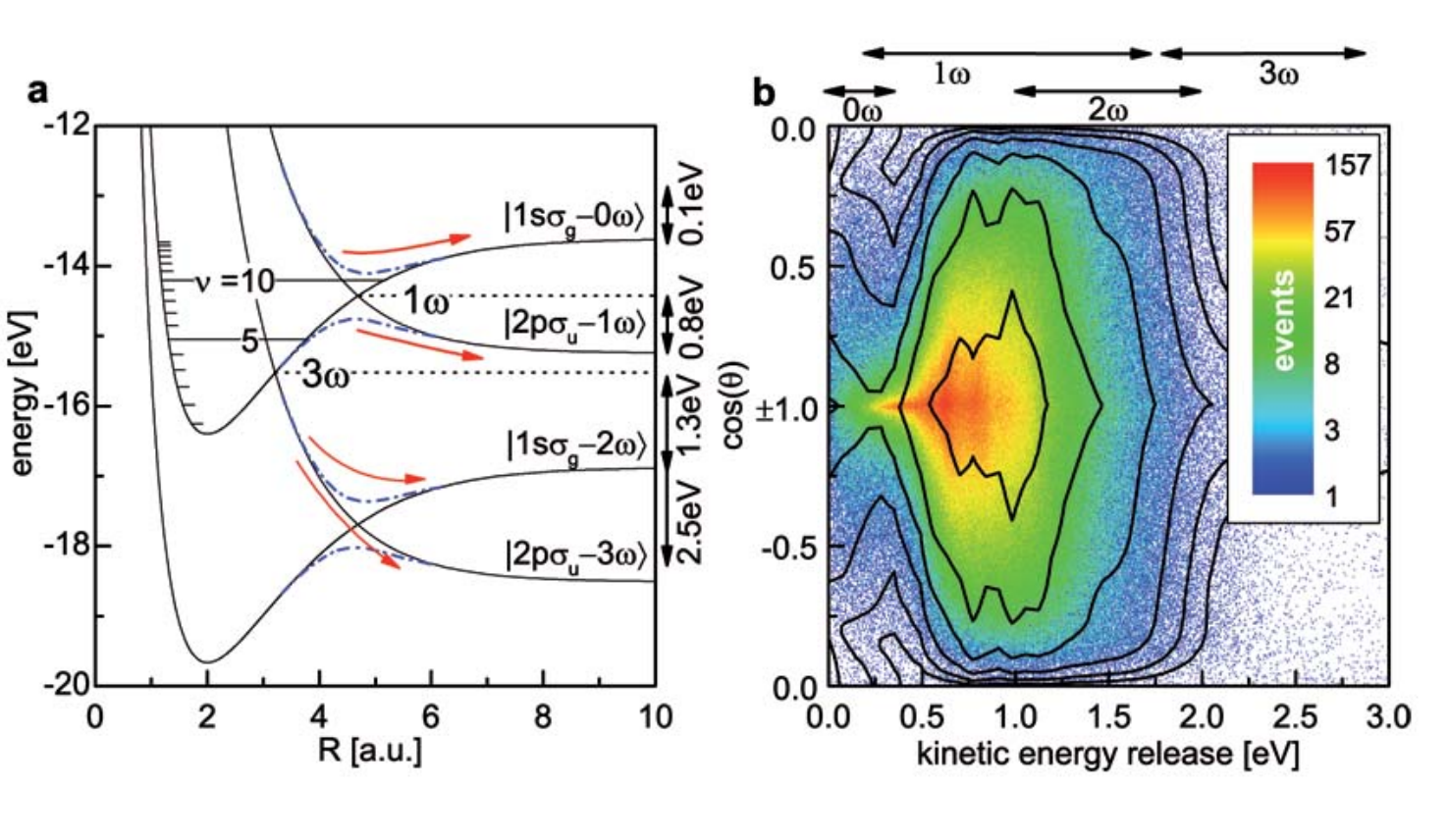}
\end{center}
 \caption{\textbf{a}, Scheme of the possible pathways of photodissociation with Born-Oppenheimer potentials in the diabatic (solid lines) and adiabatic (dot-dashed lines) Floquet representations in the function of the inner-nuclear distance R in atomic units (a.u.). The four most important dissociation pathways for the present laser parameters are indicated by red arrows at the potential crossings. The black arrows on the right vertical axis show the resulting KER for dissociation at the crossings. \textbf{b}, Comparison of the experimental (color shading) and theoretical (contour lines) distribution of dissociation events as a function of KER and angle $\theta$ between laser polarization and molecular axis. The experimental laser peak intensity is $(4\pm2)\times10^{14}$\,W/cm$^{2}$ at a 4.5\,fs pulse duration. A $\cos(\theta)$ binning is chosen to compensate for trivial effects due to the isotropic alignment of the molecules. KER regions that can be attributed to different dissociation pathways are indicated on the upper horizontal axis.}
 \label{fig:Floquet_and_cos_KER}
\end{figure}

In Fig.~\ref{fig:Floquet_and_cos_KER}a the four dissociative pathways relevant to this experiment are marked by red arrows and explained in terms of the Floquet representation \cite{Shirley1965,Posthumus2004,Chu2004}. Each pathway corresponds to different, possibly overlapping KER ranges that depend on the initial vibrational state and on the dissociation limit \cite{Jolicard1992,Bucksbaum1990,Posthumus2000}. In Fig.~\ref{fig:Floquet_and_cos_KER}b, the yield of dissociation events is displayed in false colors as a function of the KER and the angle, $\theta$, of the molecular axis with respect to the laser polarization. The data displayed are integrated over all absolute phases. The black contour lines display the event distribution calculated from a full three-dimensional (3D) time-dependent Schr\"{o}dinger equation (TDSE) discussed below, performed at the given experimental parameters with a laser peak intensity of $1\times10^{14}$\,W/cm$^2$. The calculations include intensity averaging over the interaction region as well as nuclear vibration and rotation of the molecule (for more details, see the Supplemental Material). Thus, a direct quantitative comparison with the experiment is possible. The remarkably good agreement with the measured data proves our ability to model the laser-induced molecular dynamics involved in our experiment. The structures in the KER spectra around 0.5 to 1.25\,eV in both the measured data and the calculations arise from the spectral structure of our ultrabroadband, laser pulses and should not be confused with the well-known vibrational structure seen for dissociation of H$_{2}^{+}$ with longer laser pulses \cite{Sandig2000}.

\begin{figure}
\begin{center}
\includegraphics[width=0.45\textwidth,angle=0]{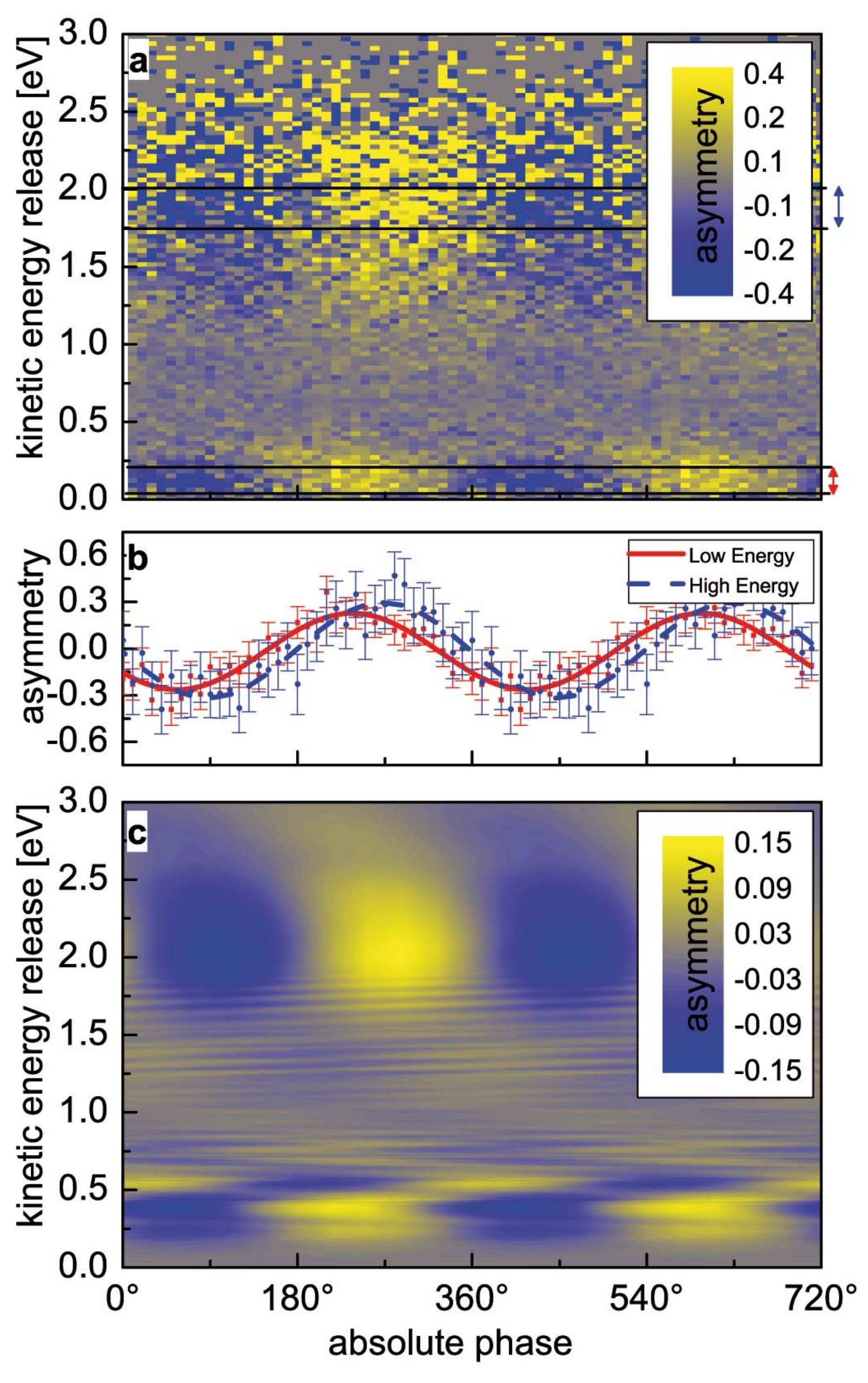}
\end{center}
 \caption{\textbf{a}, Measured asymmetry of the bound electron with a few-cycle laser pulse (4.5\,fs). \textbf{b}, Diagram of the experimental data for the two energy regions marked in a, i.e., 0.0--0.2\,eV (solid red curve) and 1.75--2.0\,eV (dashed blue curve). The curves are least-square fits of sinusoidal functions to the data points. \textbf{c}, Electron asymmetry distribution extracted from the 3D TDSE calculation. $\phi = 0^{\circ}$ corresponds to a cosine-like electric laser field. The experimental data are shifted in the absolute phase to fit the theoretical calculations.}
 \label{fig:exp_and_theo}
\end{figure}

The asymmetry, $A$, of the dissociation of H$_{2}^{+}$ is shown in Fig.~\ref{fig:exp_and_theo}a for the same data set as displayed in Fig~\ref{fig:Floquet_and_cos_KER}b. $A$ is defined as $A$=$(H_{\rm up}$$-$$H_{\rm down})/(H_{\rm up}$+$H_{\rm down})$, where $H_{\rm up}$=$H$+$p$ and $H_{\rm down}$=$p$+$H$, are the yields for dissociation with $H$ fragmenting in the upward direction and $p$ in the downward direction and vice versa, respectively. The upward direction corresponds to the electric field direction for $t$=$0$ and $\phi$=$0$. For example, $A$ is positive if more H atoms dissociate in the $\vec{E}(t$=$0,\phi$=$0)$ direction than in the $-\vec{E}(t$=$0,\phi$=$0)$ direction. The alignment angle of the molecules is restricted to within $25^{\circ}$ of the laser polarization.

In the H$_{2}^{+}$ asymmetry map, two regions with high asymmetries can be seen: $24\%\pm1\%$ between a KER of 0.0 and 0.25\,eV [low energy, LE] and 30\,$\%\pm1\%$ at 1.75 to 2.0\,eV [high energy, HE]. Fitting a $\cos(\phi-\delta)$ dependence to these regions \cite{Hua2009}, we find a difference in phase offsets $\delta_{\rm HE}$$-$$\delta_{\rm LE}$=$32^{\circ}\pm3^{\circ}$ (see Fig.~\ref{fig:exp_and_theo}b). Between these two regions the asymmetry is negligible. Comparison with the TDSE calculation again yields an excellent agreement [see Figs.~\ref{fig:exp_and_theo}a and \ref{fig:exp_and_theo}c].

It should be noted that there are small differences between experiment and theory. The most noticeable difference can be found in the amplitudes of the asymmetry. Additionally, theory exhibits a stronger energy-dependent phase shift between 0.3 and 0.7 eV than is measured. These small quantitative disagreements are likely due to the slightly lower intensity used in the calculation. This was unavoidable as a dissociation yield sufficient to elucidate the asymmetry could only be achieved at intensities where ionization -- for which the TDSE calculations fail -- was not entirely negligible \cite{Anis2008}.

Remarkably, the asymmetry maps are significantly different from those obtained with neutral hydrogen or deuterium and similar laser parameters. For example, in the KER regions of interest for dissociation (KER$<3.0$\,eV), the neutral maps show no effects \cite{Kling2006a} or a smooth diagonal shift of the localization for different KERs \cite{Kremer2009}. Obviously, control of the electron movement starting with a stationary H$_{2}^{+}$ state is fundamentally different from the dissociation-ionization process in H$_{2}$. 

\begin{figure}
\begin{center}
\includegraphics[width=0.3\textheight,angle=0]{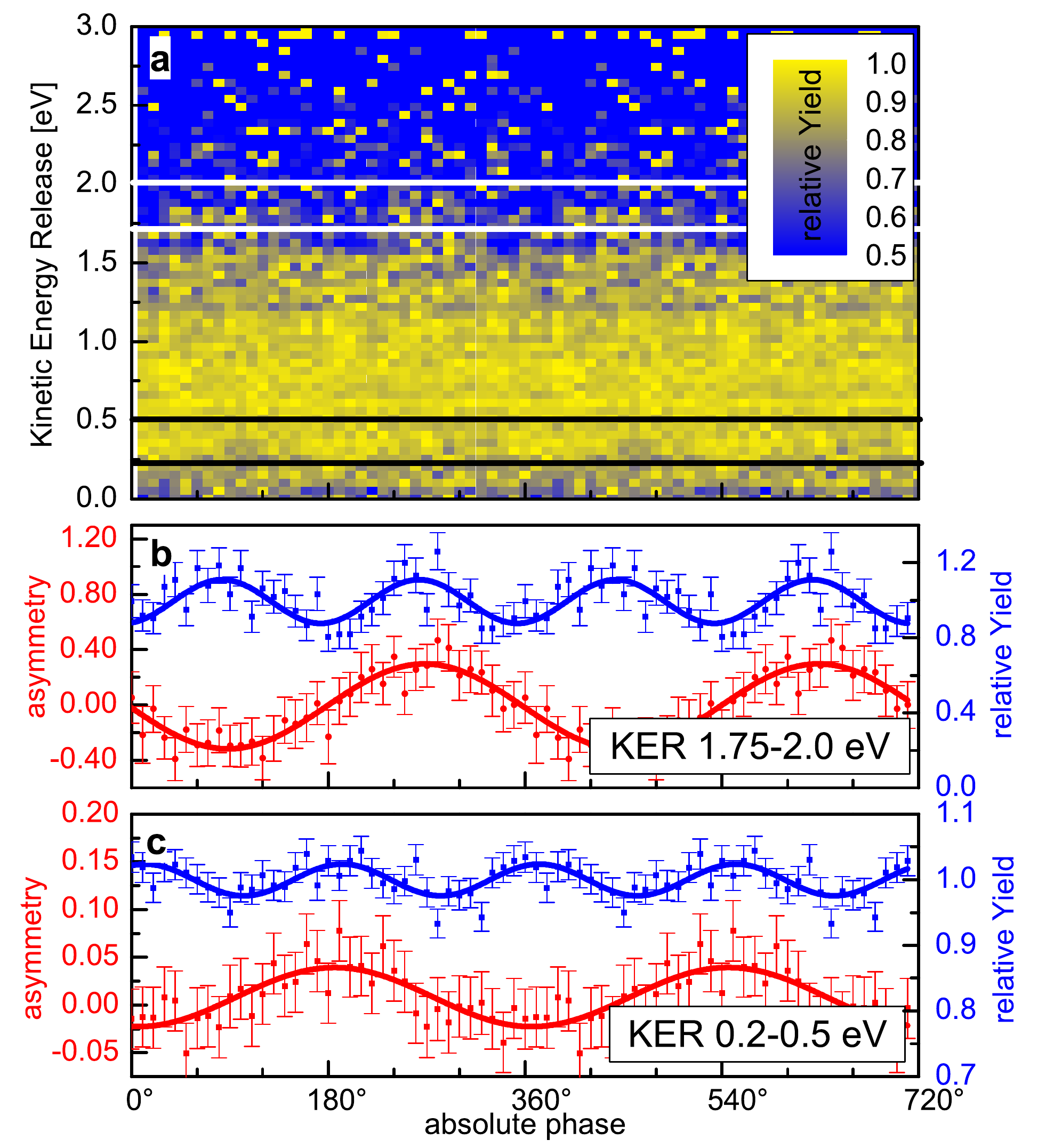}
\end{center}
 \caption{\textbf{a}, Controlling the H$_{2}^{+}$ dissociation yield with the absolute phase. The number of events in each pixel is normalized to the maximum number of events in that row, i.e., at that KER value. \textbf{b,c}, Diagram of the yield (upper blue curve) and asymmetry (lower red curve) for a KER of 0.2-0.5\,eV and 1.75-2.0\,eV. The electron asymmetry data are taken from Fig.~\ref{fig:exp_and_theo}\textbf{a}. The dissociation yield oscillates with double the frequency of the asymmetry.}
\label{fig:Yield_exp}
\end{figure}

The dissociation yield of molecules aligned within $25^{\circ}$ of the polarization is also controlled by the absolute phase and exhibits a clear $\sin(2\phi)$ modulation \cite{Hua2009}. This is evident in the KER regions between 1.75 and 2.0\,eV (with an amplitude of $11\,\%\pm1\,\%$) and between 0.2 and 0.5\,eV (with an amplitude of $2.5\,\%\pm0.3\,\%$), see Figs.~\ref{fig:Yield_exp}b and \ref{fig:Yield_exp}c. 

Control of electron localization can also be explained with a simple theoretical model. The fundamental insight in this respect is that the final state has to be a superposition of an even $|1s\sigma_{g}\rangle$ and an odd $|2p\sigma_{u}\rangle$ electronic state \cite{Cormier1998,Roudnev2007} (see the Supplemental Material). For example, a wave packet can split and propagate along both the $|1s\sigma_{g}-2\omega\rangle$ and $|2p\sigma_{u}-1\omega\rangle$ pathways as shown in Fig.~\ref{fig:Floquet_and_cos_KER}a. Both parts interfere and produce the asymmetries seen at 1.5\,eV in Fig.~\ref{fig:exp_and_theo}a. The same is true for the $|2p\sigma_{u}-1\omega\rangle$ and $|1s\sigma_{g}-0\omega\rangle$  states at 0.25\,eV. The predictions of this simple picture are in agreement with the 3D TDSE calculations, which show asymmetry effects in energy regions where the probability to dissociate along the $|1s\sigma_{g}\rangle$ and $|2p\sigma_{u}\rangle$ states are comparable in strength and the incoherent sum of vibrational states does not wash out the effect (see the Supplemental Material). 

In contrast to the asymmetry, the dissociation yield oscillates with $2\phi$ absolute phase dependence. It should be noted that the yield becomes dependent on the phase only if an interference between two pathways with the same initial state and final momentum occurs between pathways with the \textit{same} final parity \cite{Hua2009}. For example, the interference between the $|2p\sigma_{u}-1\omega\rangle$ and $|2p\sigma_{u}-3\omega\rangle$ states generates the absolute phase-dependent yield seen at 1.7\,eV in Fig.~\ref{fig:Yield_exp}a.

\begin{figure}
\begin{center}
\includegraphics[width=0.5\textwidth,angle=0]{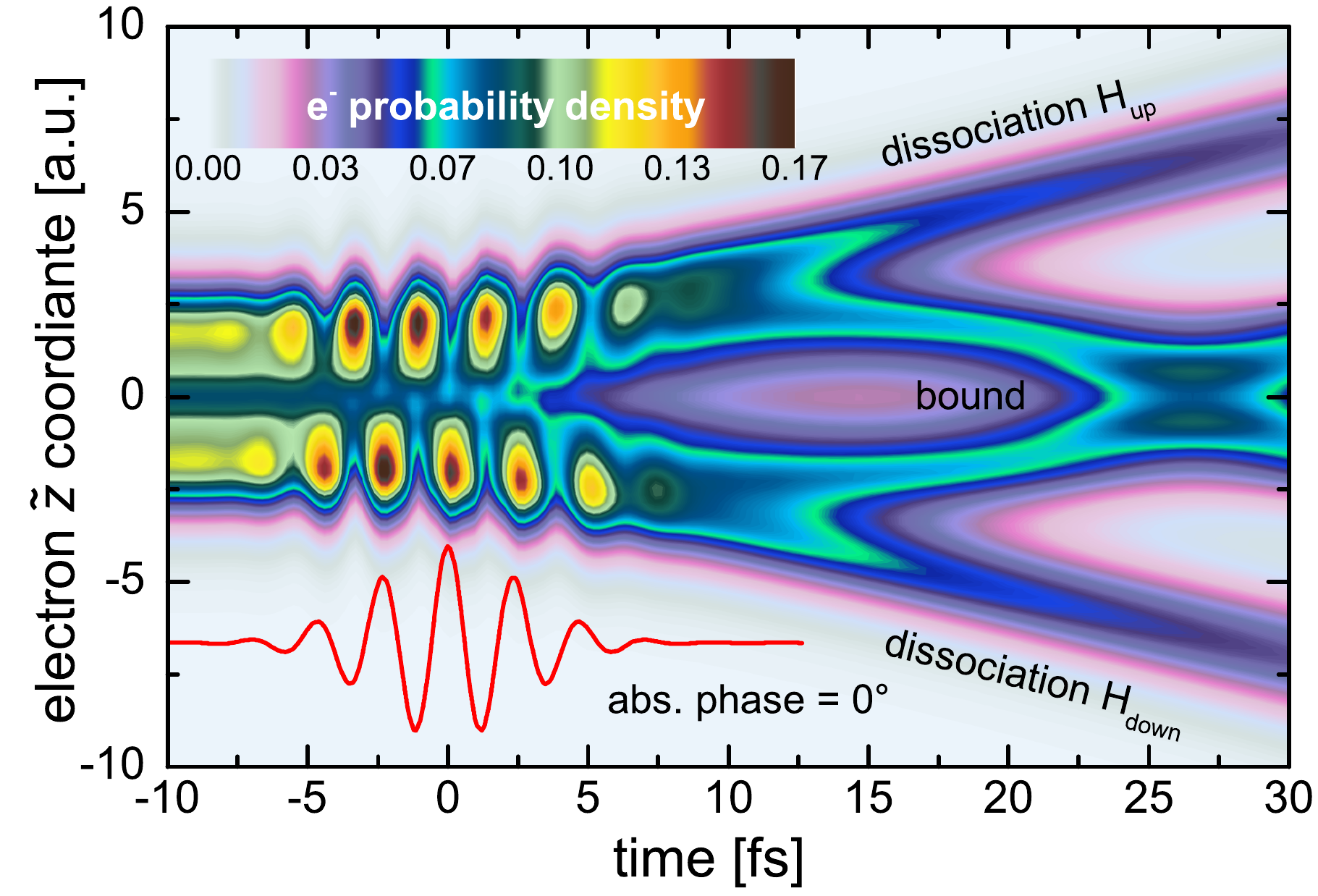}
\end{center}
 \caption{Visualization of the electronic probability density $\rho(\tilde{z},t)$ in the molecular frame $\tilde{z}$ (in atomic units) defined as $\rho(\tilde{z},t) =\int_0^{\pi/2} \sin\theta d\theta \int_0^{2\pi} d\phi  \int_0^\infty dR \int d\tilde{x}d\tilde{y} |\Psi({\bf R},{\bf r},t)|^2 $, where $\tilde{x},\tilde{y}$ and $\tilde{z}$ are electronic coordinates in the molecular frame (see the Supplemental Material for more details). The total wave function $\Psi$ is obtained from a TDSE calculation for a 4.5\,fs Gaussian pulse with the peak intensity of $1\times10^{14}$\,W/cm$^2$ (red curve at the bottom of the figure) interacting with the initial state $v=8$. }
\label{fig:3Dsimulation}
\end{figure}

In addition to quantitatively predicting the measured results, the TDSE calculations also yield the time-dependent electron motion, which gives a clear picture of the underlying attosecond dynamics involved in the dissociation process, see Fig.~\ref{fig:3Dsimulation}. Here the electron probability density is first driven by the electric field and then continues to be transferred back and forth between the nuclei until this transfer is suppressed at around 15\,fs by the increasing potential barrier and distance between the nuclei. Part of the nuclear wave packet proceeds on the dissociation path, while the rest remains bound as marked. The dissociating wave packet has a KER spread which is seen as an angular spread in this view while the bound wave packet begins to oscillate in the well.

The measurements were repeated with D$_{2}^{+}$ under similar laser parameters. 
Qualitatively, the two isotopologues display the same energy dependence since the dynamics are dictated by the Born-Oppenheimer potentials as discussed above.  Although the laser pulse is effectively shorter for D$_{2}^{+}$ \cite{McKenna2012}, the amplitudes of the asymmetry in the two aforementioned energy regions were decreased for the heavier mass by a factor of 3.5, which is in quantitative agreement with our calculations and those done previously \cite{Hua2009}. The modulation in the total yield also decreases by a factor of 3 for both energy regions as compared to H$_{2}^{+}$ , which agrees with the calculations done using the non-Gaussian laser spectrum realized in the experiment. However, these results are in a disagreement with previous calculations for a Gaussian laser spectrum \cite{Hua2009}. These quantitative discrepancies suggest a complex dependence and interplay between parameters producing absolute phase dependencies, e.g. vibrational spacing, spectral shape and mass. In addition, the magnitude of the phase offset in the asymmetry between the two energy regions increases to $\delta_{\rm HE}$$-$$\delta_{\rm LE}$$=$$-109^{\circ}\pm5^{\circ}$. 

In summary, we have realized what is one of the simplest scenarios of a photochemical reaction, namely dissociation of H$_{2}^{+}$ with virtually a single optical cycle. Further, by removing the complicating prerequisite ionization step (H$_{2}\rightarrow$H$^+_{2}$), we can more readily measure, interpret, and control the absolute phase-dependent electron motion and dissociation probability for the simplest molecule, thereby providing a benchmark for the understanding and implementation of coherent control of chemical reactions. The measurements can indeed be accurately reproduced by \textit{ab initio} calculations and differ significantly from previous measurements starting with neutral H$_{2}$ or D$_{2}$ molecules under similar conditions. Moreover, the data can be qualitatively explained by simple arguments based on the parity of the involved states. 

We thank D. Hoff for technical support. This work was realized with the Grant No. PA730/4 from the German Research Foundation (DFG) and within the Transregio 18 (DFG) and supported by the Chemical Sciences, Geosciences, and Biosciences Division, Office of Basic Energy Sciences, Office of Science, U.S. Department of Energy.



\end{document}